\documentclass{PoS}
\usepackage{url}

\title{Novel Physics with Tensor Polarized Deuteron Targets}

\ShortTitle{Novel Physics with Tensor Polarized Deuteron Targets}

\author{\speaker{Karl Slifer}\\
        University of New Hampshire\\
        E-mail: \email{karl.slifer@unh.edu}}

\author{Ellie Long\\
        University of New Hampshire\\
        E-mail: \email{ellie@jlab.org}}

\abstract{
The Jefferson Lab PAC recently approved an experiment  which will use an enhanced tensor polarized solid target.
This exciting development holds the potential of initiating a new field of tensor spin physics at JLab.
Experiments which utilize tensor polarized targets can help clarify  how nuclear properties arise from partonic degrees of freedom, provide unique insight into short range correlations and quark angular momentum, and help pin down the polarization of the quark sea.
}

\FullConference{XVth International Workshop on Polarized Sources, Targets, and Polarimetry,\\
		September 9-13, 2013\\
		Charlottesville, Virginia, USA}

\begin{document}

\section{Background}
\begin{figure*} 
        \begin{center}
                \includegraphics[width=6cm]{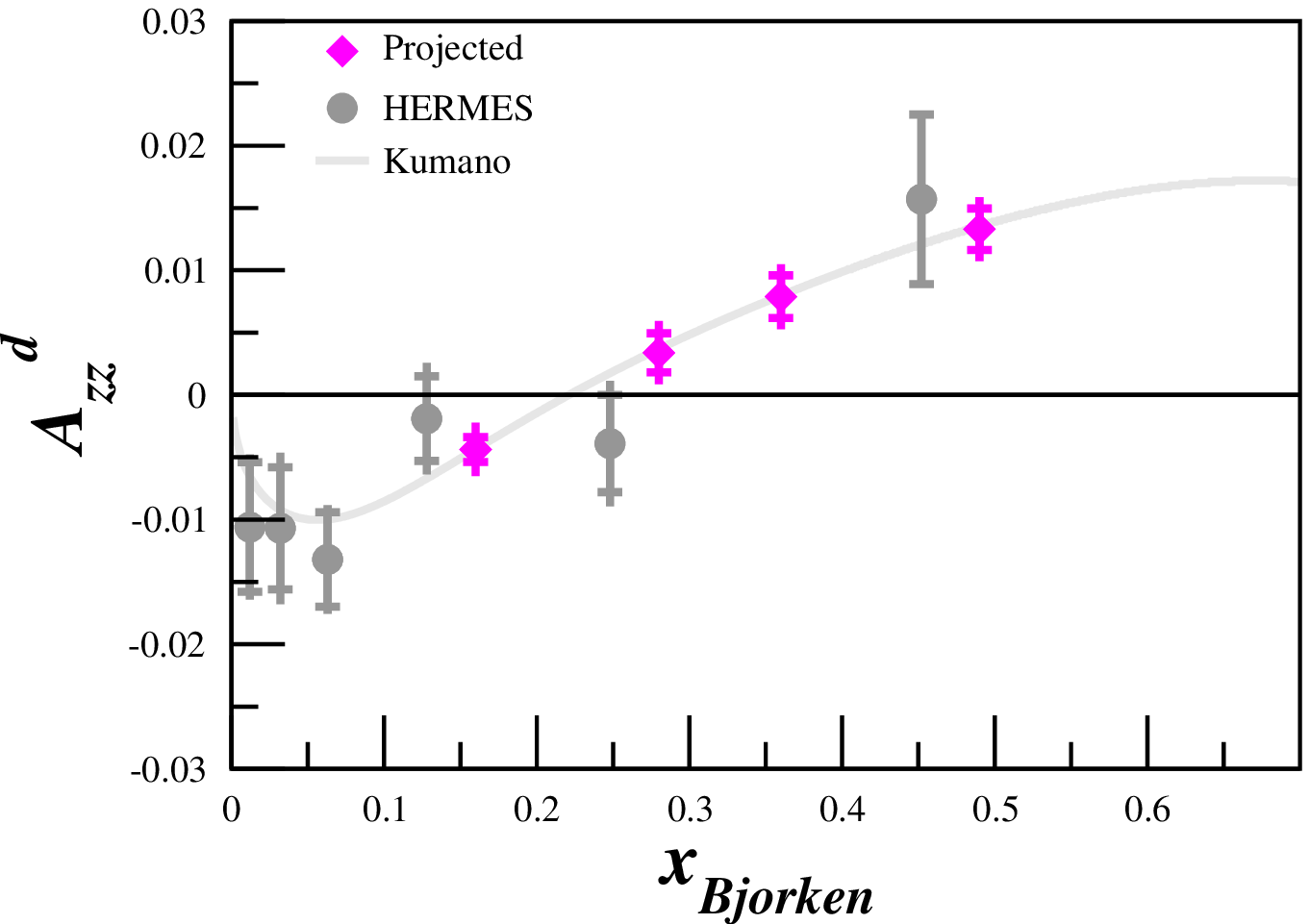}
                \hspace{1cm}
                \includegraphics[width=6cm]{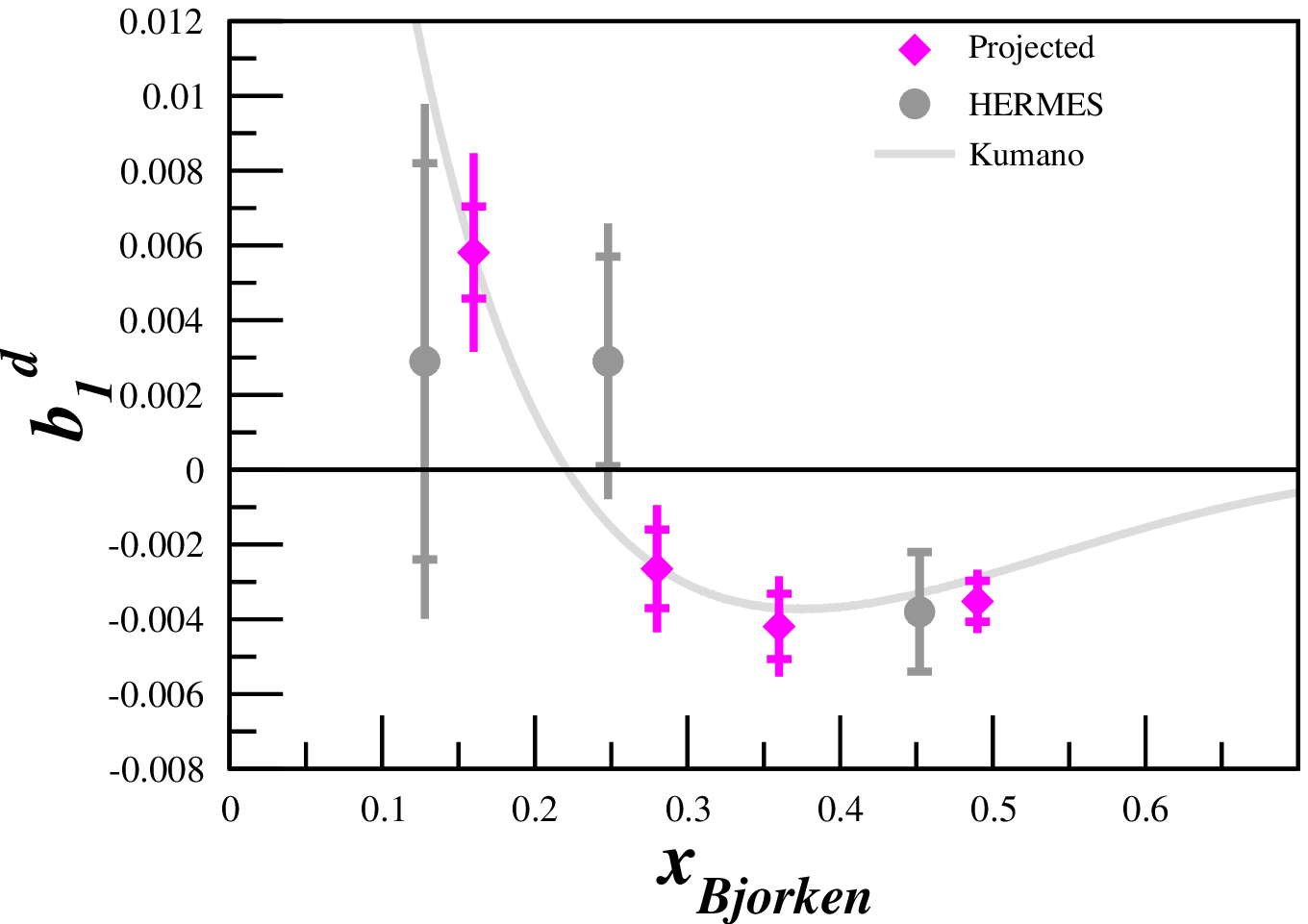}
                \caption[Proposed Measurements and Estimated Uncertainties of $A_{zz}$ and $b_1$]{Proposed measurements and estimated uncertainties of $A_{zz}$ and $b_1$ assuming a tensor-polarized deuterated ammonia target capable of reaching $35\%$ polarization. The estimates are plotted alongside previous data from HERMES \cite{Airapetian:2005cb} and a fit from Kumano \cite{Kumano:2010vz}.}
                \label{fig:azz-b1}
        \end{center}
\end{figure*}

The Jefferson Lab PAC 40 recently approved the E12-13-011 experiment~\cite{Slifer:2013} to measure the leading-twist tensor structure function, $b_1$, of the deuteron. This observable provides unique insight into how coherent nuclear properties arise from partonic degrees of freedom in the simplest nuclear system.
At low $x$, shadowing effects are expected to dominate $b_1$, while at larger values, $b_1$ provides a clean probe of exotic QCD effects, such as hidden color due to a 6-quark configuration. Since the deuteron wave function is relatively well known, any novel effects are expected to be readily observable. 
There are several different approaches for predicting the $x-$dependent behavior of $b_1$, including one pion exchange~\cite{Miller:1989nc}, convolution models with relativistic and binding energy corrections~\cite{Khan:1991qk}, covariant relativistic calculations~\cite{Umnikov:1996qv}, double-scattering effects~\cite{Bora:1997pi}, and the virtual nucleon approximation~\cite{MISAK}.

All of these models predict a small or vanishing value at moderate $x$. However, the first measurement of $b_1$ at HERMES revealed a cross-over to a large negative value in the region $0.2<x<0.5$, albeit with relatively large experimental uncertainty. Kumano~\cite{Kumano:2010vz} points out that the twist-2 structure functions $b_1$ and $b_2$ can be used to probe orbital angular momentum.  He then extracted the tensor polarized quark and anti-quark distributions from a fit to the HERMES data~\cite{Airapetian:2005cb} and found that a non-negligible tensor polarization of the sea is necessary to reproduce the trend of
the data. 

E12-13-011 will perform an inclusive measurement of the deuteron tensor asymmetry in the
region $0.16<x<0.49$, for $0.8<Q^2<5.0$ GeV$^2$.
The UVa/JLab solid polarized ND$_3$ target will be used, along with the
Hall C spectrometers, and an unpolarized  beam.
This measurement  will provide access to the tensor quark polarization, and provide  data needed to test the
Close-Kumano sum rule, which has been predicted to be satisfied~\cite{Close:1990zw} in the absence of tensor polarization in the quark
sea.

The PAC approval is conditional on demonstration of $30\%$ tensor polarization in a solid polarized target.

\subsection{Potential New Measurements}
The approval of E12-13-011 has already stimulated new treatments of tensor degrees of freedom in nuclei.  In particular, both Miller~\cite{Miller:2013hla}, and Sargsian~\cite{MISAK} have new predictions for the tensor asymmetry in the DIS and $x>1$ region, respectively.  The new experiment has also spurred a renewed discussion of the validity of the Close-Kumano Sum Rule~\cite{Close:1990zw}. 

The approval of E12-13-011 opens the possibility of pursuing several interesting new experiments at JLab.
 We discuss here a few of the most promising potential measurements that could be performed  with a tensor polarized target.

\subsubsection*{The Tensor Asymmetry $A_{zz}$ from D($e,e'$) in the Quasi-Elastic and $x>1$ Region}
The tensor asymmetry, which is used to extract $b_1$ in the DIS region through the D($e,e'$)X channel, has also been calculated in the quasi-elastic $x$-region; First by Frankfurt and Strikman, using the Hamada-Johnstone and Reid soft-core wave functions~\cite{Frankfurt:1988nt}, and more recently by {Sargsian} for $x>1$ using a virtual-nucleon as well as a light-cone approach~\cite{MISAK}. In the $x>1.4$ plateau region, the two approaches differ by approximately a factor of 2. 

These asymmetries are much larger than expected for the $b_1$ measurement, which simplifies many of the technical challenges that needed to be addressed in the already approved proposal. Our initial estimates indicate that the tensor asymmetry in the previously unexplored large Bjorken x region can be mapped out with 9\% total relative uncertainty with a two week run using the same installation as E12-13-011.

\subsubsection*{The Tensor Asymmetry $A_{zz}$ from D($e,e'p$) in the Low $Q^2$ Region}
The E97-102 experiment was approved by JLab's PAC13 with an A$^-$ physics rating, but was not defended during JLab's jeopardy review and was removed from the schedule.
The PAC commented that: 
\begin{quote}
``The structure of the ground state wave function in nuclei at small interparticle distance is still an unsolved problem. The deuteron is a suitable target to start this investigation because its structure can be calculated with high precision using realistic NN potentials."
\end{quote}
E97-102 would have measured $A_{zz}$ in the D($e,e'p$) channel in order to extract the cross-sections with the deuteron in the $m=\pm1$ and $m=0$ spin states. The measurement would be over the range $0.051< Q^2 < 0.185~\mathrm{GeV^2}$ and $0.2 < p_{miss} < 0.4 ~\mathrm{GeV}$, which is dominated by the $D-$state wave function of the deuteron ground state. 

\subsubsection*{The Tensor Asymmetry $A_{UT}$ and the Orbital Angular Momentum Sum Rule} 

The spin-1 angular momentum sum rule discussed by Liuti {\it et al.}~\cite{Taneja:2011sy} provides a test of whether the orbital angular momentum of the deuteron can be derived directly from quark contributions of individual protons and neutrons, or whether there are other contributions,  which must be accounted for, such as from gluons.
By examining the energy momentum tensor for the deuteron, the authors showed that
it was possible to define an additional sum rule 
where it was demonstrated that the second moment of this quantity is non vanishing, being related to one of the gravitomagnetic deuteron form factors.

To study this sum rule, the authors suggest a measurement of the tensor-polarized deuteron's transverse spin asymmetry, $A_{UT}$, which is derived in terms of GPDs~\cite{Berger:2001zb,Berger:2001xd}. It is also important to notice that $b_1$ singles out the role of the $D$-wave component in distinguishing coherent nuclear effects through tensor polarized correlations  from  the independent nucleon's partonic spin structure.
A similar role of the $D$-wave component was also found in the recently proposed spin sum rule where it plays a non-trivial role
producing a most striking effect through the spin flip GPD, $E$.

\subsubsection*{The Tensor Structure Functions $b_2$, $b_3$, and $b_4$}
The leading-twist structure function $b_1$ is only one of four tensor structure functions needed to describe inclusive scattering from a spin-1 hadron. The second tensor structure function  $b_2$ is expected to be related to $b_1$ through a Callan-Gross-type relation, and the higher twist $b_3$ and $b_4$ have not been directly explored experimentally.

\subsubsection*{The Tensor Observable $T_{20}$}
The tensor observable $T_{20}$ was studied previously at Jefferson Lab \cite{PhysRevLett.82.1379,Abbott:2000ak} and elsewhere \cite{PhysRevLett.52.597,Dmitriev1985143,Gilman:1990vg,Boden:1990una,Garcon:1989gy, The:1991eg,Garcon:1993vm}.  Together with $A(Q^2)$ and $B(Q^2)$,  the tensor polarization moment $T_{20}$ is used to
extract the three form factors of the spin-1 deuteron: the charge monopole $G_C$, the charge quadruple $G_Q$, and the magnetic dipole $G_M$.
These form factors can be extracted from cross-section and polarization moment measurements through electron scattering on tensor-polarized deuterium.
In the one-photon exchange approximation, the deuteron cross-section is defined by the Mott cross-section and a linear combination of $A(Q^2)$ and $B(Q^2)$. At low $Q^2$, $T_{20}$ is dependent only on the ratio $G_Q/G_C$. Combined with $A(Q^2)$'s dependence on all three form factors and $B(Q^2)$'s on only $G_M$, a measurement of all three quantities can be used to extract each of the form factors.  
There have periodically been calls to extend the existing measurements to large $Q^2$, but a compelling physics motivation for these new measurements needs to developed.

\subsection{The Tensor Polarized Target}

The Hall A/C solid polarized target operates on the principle of Dynamic Nuclear Polarization (DNP) to
enhance the low temperature, high magnetic field polarization of solid
materials  by microwave pumping.
The permeable target cells are
immersed in liquid helium and maintained at 1 K by use of a
high power evaporation refrigerator.
The superconducting Helmholtz coils have a 
conically shaped aperture along the beam axis
which allow for unobstructed forward scattering.
%
The target material is exposed to microwaves
to drive the hyperfine transition which  aligns the nucleon spins.
The standard DNP technique produces  maximum (average) deuteron vector polarizations of up to 60\% (30\%) in ND$_3$,
which corresponds to maximum (average) tensor polarizations of approximately 35\% (9\%), respectively.


%
Tensor polarization can be significantly enhanced from these baseline values by disturbing the thermal equilibrium
of the sample using a frequency modulated RF source to stimulate transitions from the $m=0$ level.
This method of `burning holes' in the NMR line with a saturating RF field was demonstrated by deBoer~\cite{deBoer:1974vr}, Meyer~\cite{Meyer:1985ju,Meyer:1985dta}, and Bueltmann {\it et al.}~\cite{CRABBLAB}.

%
A second possible method to simultaneously stimulate transitions from two of the
deuteron's three sublevels uses independent microwave sources tuned to the closely spaced ESR transitions.
Initial unpublished studies~\cite{CRABBLAB} indicated that this technique is feasible, with
the caveat that care must be taken to prevent resonant coupling in the two microwave cavities.


%

The existing target underwent significant renovation during the recent E08-027 run~\cite{g2p}.  Most notably, the target Helmholtz coils quenched catastrophically early in the run, and were replaced with coils
scavenged from the decommissioned Hall B target.  The 12 GeV polarized target program will benefit if this target can be replaced or supplemented with a device optimized for the next round of experiments.

%


\section{Summary}
We anticipate that the development of tensor polarized solid targets will motivate an exciting new generation of spin-dependent measurements at Jefferson Lab.  This process has begun with the recent approval of the E12-13-011 experiment, and there is an robust research program that awaits development for future investigations.

\end{document}